# Large Superconducting Magnet Systems

*P. Védrine[1]*
CEA Saclay, Gif sur Yvette Cedex, France

**Abstract**
The increase of energy in accelerators over the past decades has led to the design of superconducting magnets for both accelerators and the associated detectors. The use of Nb–Ti superconducting materials allows an increase in the dipole field by up to 10 T compared with the maximum field of 2 T in a conventional magnet. The field bending of the particles in the detectors and generated by the magnets can also be increased. New materials, such as $Nb_3Sn$ and high temperature superconductor (HTS) conductors, can open the way to higher fields, in the range 13–20 T. The latest generations of fusion machines producing hot plasma also use large superconducting magnet systems.

## 1    Overview of large-scale applications of superconductivity

Nowadays, superconductivity is essential in the construction of magnets for physics. By using superconducting magnets, the Joule effect can be eliminated, because the electrical resistance below a critical temperature is zero. The magnetic field generated in very large volumes can also be increased compared with that in conventional magnets. The operating cost of the machine can be reduced, despite the additional equipment needed to maintain the magnet at low temperature, and the size of the magnets is decreased because of the high current density in the superconducting materials.

The materials used in the design of superconducting magnets for physics are metallic alloys such as Nb–Ti or inter-metallic compounds such as $Nb_3Sn$. They can carry current densities as high as 3000 $A \cdot mm^{-2}$ at 12 T and 4.2 K.

In particle physics accelerators, the increase in the collision energy leads to higher bending fields and higher focusing gradients, because the particle energy of a beam is proportional to its radius multiplied by the dipole bending field. A normal conducting magnet design is limited both to fields below 2 T and to a gradient below 20 $T \cdot m^{-1}$, which would lead to a very large synchrotron. The first large accelerator built with superconducting magnets was the Tevatron at the Fermi National Laboratory near Chicago in 1983 [1]. The recent discovery of the Higgs boson at the Large Hadron Collider (LHC) highlights the role of superconductivity in this machine, which has been in operation at CERN since 2008 [2].

Superconducting magnets are used not only in the accelerator, but also in the associated detectors. The first superconducting magnet in a detector came with the 10 in. bubble chamber magnet built at the Argonne laboratory (1964) [3]. The most popular design for detectors is a solenoid with diameters of a few metres and a length of tens of metres. However, dipole or toroidal configurations have also been built for some detectors. The ATLAS detector built for the LHC is one of the most famous examples of a toroidal magnet [4].

---

[1] pierre.vedrine@cea.fr

Controlled thermonuclear fusion is also a very interesting large-scale application and requires magnetic confinement of plasma with very high fields (>13 T) in large volumes. Several machines have been built in the past and remain in operation. The superconducting magnets for the International Thermonuclear Experimental Reactor Programme (ITER) built in Cadarache, France, are now under construction [5].

Superconducting magnets are also used in many areas other than high-energy physics or fusion. Superconducting materials must be used everywhere that high fields in large volumes are required, or where increased current densities or a reduction of the Joule effect is desired. Superconducting solenoids are now a common tool in research laboratories, providing fields from a few tesla to more than 20 T in bores from 30 to 300 mm. Superconducting magnets for magnetic resonance imaging are used commonly in hospitals, and new research projects are pushing the limits of image resolution [6]. Nuclear magnetic resonance spectroscopy systems used to obtain information about molecules rely on high-field superconducting magnets. Other applications, such as power applications (rotating electrical machines, transformers, fault current limiters, power transmission lines), transportation (ship and aircraft propulsion systems, magnetically levitated trains, railway traction transformers), industrial applications (inductive heating, magnetic separation), and energy storage are all under investigation [7].

The market for superconductivity, for an annual total of €5 million, is today dominated by the magnetic resonance imaging market, and this is expected to increase slowly during the next decade. Applications with high-temperature superconductors represent less than 1% of the total.

## 2 Accelerator magnets: review of the four major projects (Tevatron, HERA, RHIC, and LHC)

### 2.1 Introduction

The challenge in manufacturing superconducting magnets for accelerators is to ensure reproducibility, safety, and reliability of hundreds of magnets mounted along kilometres of a circular accelerator, in a cryogenic environment.

The design of superconducting magnets is very different from that of conventional magnets. The quality of the magnetic field is obtained principally from the position of the conductors around the bore tube, and the tolerances on the positioning must be better than 20 $\mu$m. An iron yoke is present to reduce the fringe field outside the coil and also to enhance the field in the useful aperture of the magnet. Some specific problems arise in the use of superconducting magnets, for example quenches, training, and protection.

The Lorentz forces, created by the combination of the high current density in the coils and the flux density, tend to squeeze the coils azimuthally towards the mid-plane with a magnitude of many 100 kN·m$^{-1}$. These forces have the tendency to move the coil from its end stops at the pole plane. In the same way, the radial forces tend to force the coil outwards with a maximum displacement at the mid-plane. Axially, the coils tend to elongate. All these forces have to be maintained in a strong mechanical structure to avoid deformation or movement of the coils, which would lead to premature quenches or a bad field quality.

One way to prevent movement of the conductors under the azimuthal forces is to apply a pre-compressive force during the clamping of the collars, which maintains the coils in a rigid cavity. The coils act like springs with a non-linear hysteretic response, and the Lorentz forces have to pass beyond the pre-load value to begin to displace the conductors significantly. The value of the pre-compression required at room temperature depends on the mechanical arrangement and on the relative thermal contraction of the selected materials between the room and helium temperatures.

Generally, the coil consists of two layers of turns of Rutherford cables, sometimes different between one layer and another. The cable is wound with a tension of about 500 N on a mandrel assembled with accurately stamped laminations. Once the first layer is ready, it is cured under pressure in a mould also made with stamped laminations. This operation is performed at the curing temperature of the resin contained in the insulation of the cable. The second layer of the coil is wound separately on a second mandrel or on top of the first layer after its curing. The second solution has the advantage of avoiding a spliced joint between the layers.

After the curing of the second layer, the coil is ready to be assembled with the other coils in common collars (punched laminations a few millimetres thick), made of aluminium alloy or austenitic stainless steel. The clamping of the collars is insured by rods or keys. The dimensions of the collars are set to ensure the pre-load required for the magnetic forces and the correct geometrical shape of the coil.

The stacking of the iron laminations can take place around the collared coil. The iron yoke is generally compressed onto the collars and fixed by clamps. The gap between the top and bottom (or left and right) iron laminations is open at room temperature and designed to be closed at helium temperature.

An outer cylinder in one or two parts, made of stainless steel or Al alloy, is welded or slid longitudinally around the yoke to align the entire assembly and to provide an additional pre-load on the coil.

If no precautions have been taken, when a magnet quenches the entire stored energy will be dissipated in the area of the quench initiation point. In high-field accelerator magnets, the need for a high overall current density implies a low copper to superconducting ratio (about 2:1). We are therefore very far from cryogenic stabilization, and most of the heat dissipated in the normal zone will heat the quench initiation point. The only way to decrease the temperature is to obtain a faster current drop by spreading the quench rapidly over the entire coil in which the stored energy can be absorbed.

## 2.2 Tevatron and HERA

The first complete superconducting synchrotron was the Tevatron built at FNAL near Chicago [8]. The Tevatron was commissioned in 1983 and closed in 2011; during operation in 1995 it was used to discover the top quark. The 1 TeV proton–antiproton ring was made of about 774 superconducting dipoles and 216 quadrupoles, wound with Nb–Ti conductors and assembled with a circumference of 6.3 km. The second accelerator, an electron–proton ring called HERA (Hadron-Elektron-Ring-Anlage, or 'Hadron Electron Ring Facility') with a 820 GeV proton beam, was commissioned at DESY near Hamburg (Germany) in 1991 and closed in 2007. It used about 422 dipoles and 256 quadrupoles, also over a circumference of 6.3 km [9].

Figure 1 (left) shows the Tevatron accelerator complex and (right) the HERA accelerator. Table 1 summarizes the main characteristics of the dipoles and quadrupoles of these two machines. Figure 2 shows as an example the Tevatron and HERA dipoles.

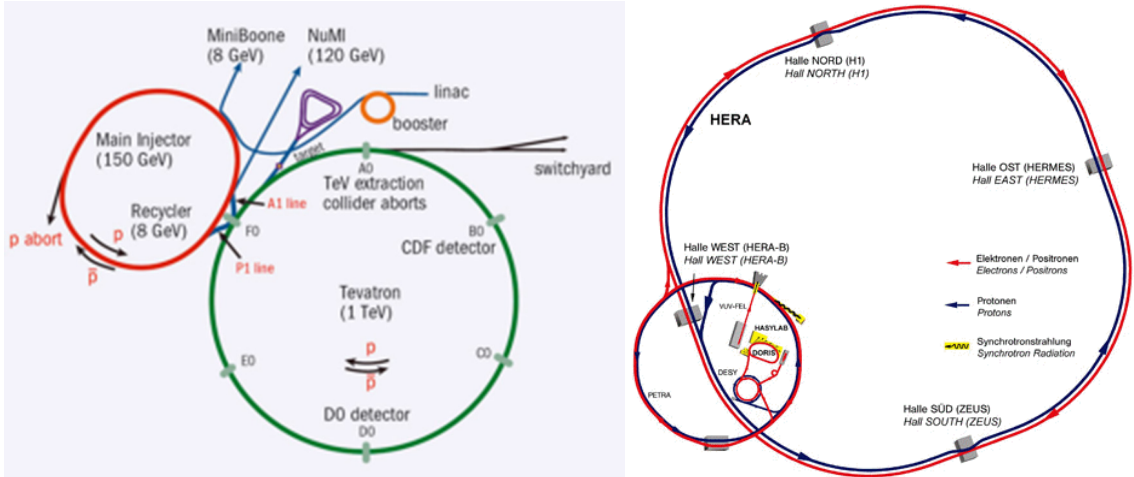

**Fig. 1:** Tevatron accelerator complex (left) and HERA accelerator (right)

**Table 1:** Magnet parameters of the Tevatron and HERA proton ring

|  | Dipole | | Quadrupole | |
| --- | --- | --- | --- | --- |
|  | **Tevatron** | **HERA** | **Tevatron** | **HERA** |
| Operating temperature (K) | 4.6 | 4.5 | 4.6 | 4.5 |
| Central field (T) | 4.4 | 4.68 | | |
| Gradient (T·m$^{-1}$) | | | 75.8 | 91.2 |
| Field length (m) | 6.11 | 8.82 | 1.679 | 1.861 |
| Nominal current (A) | 4400 | 5027 | 4400 | 5027 |
| Inner coil diameter (mm) | 76.2 | 75 | 88.9 | 75 |
| Number of coil layers | 2 | 2 | 2 | 2 |
| Superconducting material | Nb–Ti | Nb–Ti | Nb–Ti | Nb–Ti |
| Filament diameter ($\mu$m) | 8 | 14 | 8 | 19 |
| Collar material | Stainless steel | Al alloy | Stainless steel | Stainless steel |
| Yoke | Warm | Cold | Warm | Cold |
| Cryostat length (m) | 6.4 | 9.77 | 2.31 | 3.98 |
| Total number | 774 | 422 | 256 | 224 |

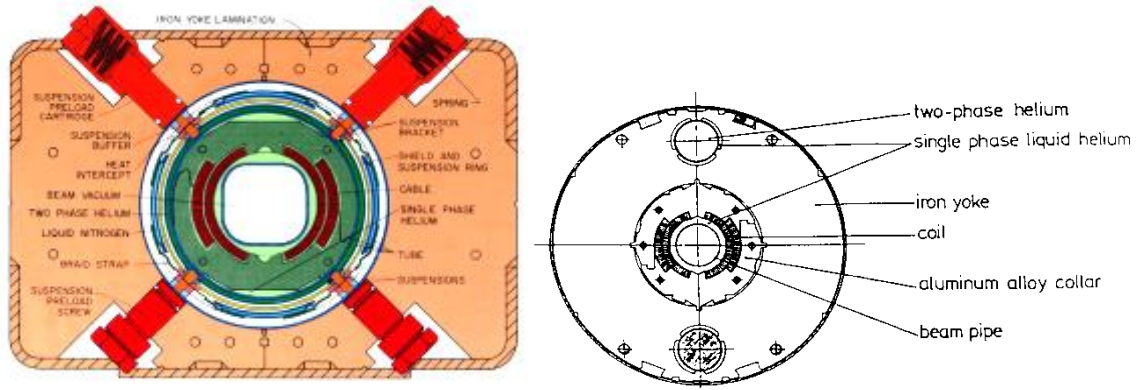

**Fig. 2:** Tevatron dipole (left) and HERA dipole (right) magnet cross-section

The superconducting magnets of both machines use Nb–Ti as the superconductor. The Tevatron used a wire with filaments of 9 $\mu$m for the dipole and the quadrupole, whereas HERA used a 14 $\mu$m filament for the dipole and a 19 $\mu$m filament for the quadrupole. The wire surfaces were quite different. The Tevatron wire surface was tinned and oxidized in alternating wires, whereas it was only silver tinned in HERA.

The cables for the Tevatron magnets were of Rutherford type, but with different numbers of strands: 23 and 24 strands of 0.648 mm cable for the dipole and quadrupole, respectively. The cross-section was 1.067/1.372 × 7.671 mm$^2$. The Cu/Sc ratio was 1.8:1. The minimum critical current density was 1800 A·mm$^{-2}$ at 4.2 K, 5 T. The insulation of the cable comprised one layer of polyimide film of 25 $\mu$m wrapped with an overlap of 58%. In addition, the cable was wrapped with fibreglass tape impregnated with curable B-stage epoxy.

The cable for HERA was a 24-strand cable with a cross-section of 1.28/1.67 × 10 mm$^2$. The Cu/Sc ratio was 1.8:1, and the minimum critical current density was 2200 A·mm$^{-2}$ at 4.2 K, 5.5 T.

The coils were wound in two layers and connected by splicing, except on the HERA quadrupole, where the second layer was wound on the first after curing.

Laminated stainless steel in the case of the Tevatron dipole, or a laminated aluminium collar for the HERA dipole, was used to clamp the coils together and apply the required pre-compression. The collars were able to withstand the magnetic forces for the Tevatron, but not for the HERA dipole, above a field of 5.9 T. At these levels of field, the yoke helped to counterbalance the forces.

The Tevatron magnets used a 'warm iron' yoke outside the cryostat (see Fig. 2), whereas the HERA magnets used a 'cold hybrid iron' yoke. In the case of the 'warm iron', the iron was at room temperature and not cooled to cryogenic temperature. The effect of the iron on the field was small because the ratio of the radii was low and the saturation negligible. The disadvantage was that the centring of the coil inside the yoke was very difficult and required cryogenic support to limit the heat conduction. A misalignment of the yoke could create very huge non-symmetric magnetic forces on the conductors, together with a field distortion. For the 'cold hybrid iron', the iron was cold, but a gap was left between the coils and the yoke where the clamping collars held the electromagnetic forces, which allowed good centring. The effect of the iron saturation was moderate, but the contribution to the main field remained important.

All Tevatron magnets were measured over two different excitation cycles: one at the nominal ramp rate and the other at double the ramp rate. Magnets quenched between 3900 A and 4900 A, with a mean value around 4500 A.

Training of the HERA magnets was negligible, and the maximum quench current was reached after only a few quenches. Measurements of the quench currents at 4.75 K gave a mean of 6458 A for the dipoles and of 7383 A for the quadrupoles.

## 2.3   The RHIC project

The Relativistic energy Heavy-Ion Collision (RHIC) project [10] is a colliding beam facility that has been in operation at Brookhaven National Laboratory since 2000. It is dedicated to studies of nuclear phenomena in RHICs at a maximal energy of 100 GeV/u. The collider is designed with two independent, identical quasi-circular concentric superconducting magnet rings (3.8 km in circumference) (Fig. 3). The two beams can intersect at six locations around the ring. A standard cell consists of two dipoles of 9.46 m and two combined magnet units comprising a quadrupole of 1.13 m, a 0.75 m sextupole, and a 0.58 m assembly of multipole correctors.

The magnets are installed in an existing tunnel. As the radius of curvature is relatively large compared with the target energy, the magnetic field of the dipole is only 3.45 T.

The dipole magnet is designed with a one-layer coil containing four blocks of conductors, separated by three wedges to improve the field quality (Fig. 4). The diameter of the bore is 80 mm. Table 2 summarizes the main parameters of the RHIC magnets.

The superconducting cable, 9.73 mm wide with a keystone angle of 1.2° (1.166 × 9.73 mm), is a 'Rutherford-type' cable with 30 strands of 0.648 mm in diameter. Each wire is a composite of 6 $\mu$m Nb–Ti filaments in a matrix of pure copper with a Cu/Sc ratio of 2.25:1. The minimum specified critical current density is 2600 A·mm$^{-2}$ at 5 T, 4.2 K, but a value of about 2800 A·mm$^{-2}$ was achieved during the manufacture of the wire. Standard polyimide/fibreglass insulation is used for the cable. The coil, which is made of 32 turns arranged in four blocks, is surrounded by a spacer made with an injection-moulded mineral-loaded phenolic glass, which acts as an insulator and creates a distance from the surrounding yoke (Fig. 4). The iron yoke serves to clamp the coils, create the pre-compression, and withstand the magnetic forces. The outer helium vessel is made with two half stainless steel tubes welded longitudinally around the yoke. The coil end forces are also retained by thick endplates anchored at the shell.

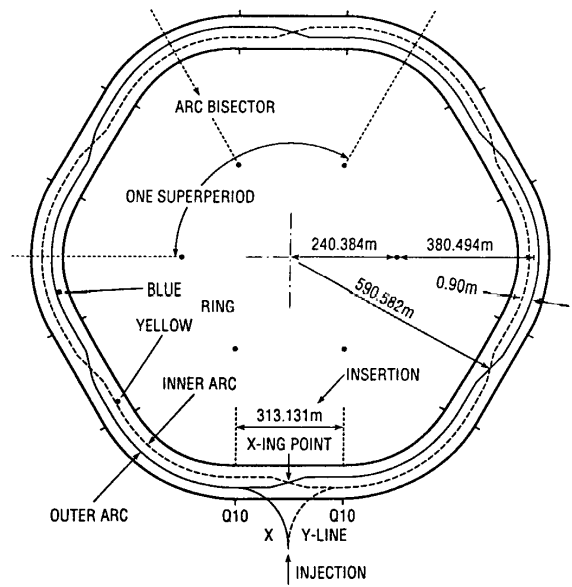

**Fig. 3:** RHIC collider

**Table 2:** Main parameters of the RHIC magnets

|  | **Arc dipole** | **Arc quadrupole** |
|---|---|---|
| Operating temperature (K) | 4.6 | 4.6 |
| Central field (T) | 3.45 |  |
| Gradient (T·m$^{-1}$) |  | 71.8 |
| Nominal current (A) | 5050 | 4720 |
| Field length (m) | 9.45 | 1.13 |
| Inner coil diameter (mm) | 80 | 80 |
| Cryostat length (m) | 9.728 | 3.023 |
| Total number | 264 | 276 |

The quadrupole magnets employ a similar design. This quadrupole is assembled with a corrector magnet at one end and a sextupole magnet at the other in one cryostat.

Of the magnets, 20% were cold tested and quenched. They reached the 'plateau' rapidly, within a few quenches. The initial quench current and the plateau quench current of each magnet exceeded the 5 kA operating current.

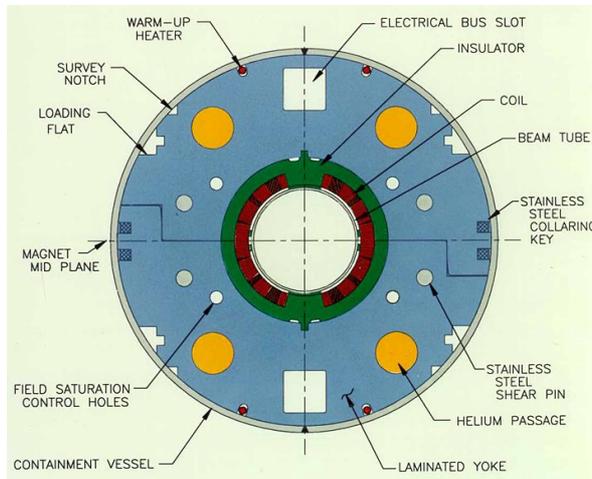

**Fig. 4**: Cross-section of RHIC dipole cold mass

## 2.4 The LHC project

The Large Hadron Collider (LHC) is a European project superconducting synchrotron, which is installed in the 27 km Large Electron–Positron (LEP) tunnel at CERN [2]. It is designed to accelerate two independent beams of protons in opposite directions up to 7 TeV, to study quark–quark interactions and to explore rare processes such as the existence of the Higgs boson. The structure of the LHC comprises eight arcs separated by eight straight sections (Fig. 5). The two beams alternate from the outer to inner channels in successive arcs, crossing each other's paths at eight points, six of which could be used for physics experiments. The circumference of the collider is fixed by the dimensions of the LEP tunnel, in which available space is limited.

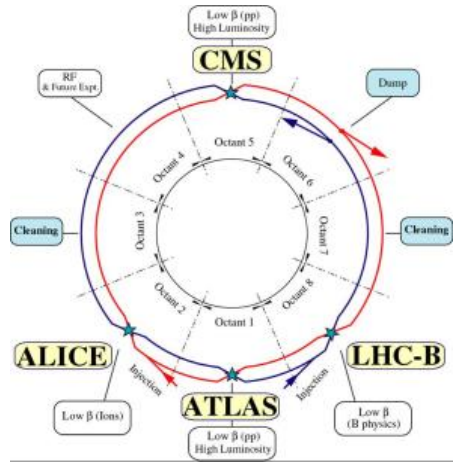
**Fig. 5:** LHC collider

These two constraints have led to the development of the concept of a high-field superconducting magnet of 8.33 T assembled in a structure referred to as 'two-in-one'. The two magnetic apertures are placed horizontally in the same yoke and cryostat, leading to a smaller cross-section (Fig. 6). The choice of Nb–Ti as the superconductor has also led to the use of superfluid helium to reach the high field demanded. The use of superfluid helium to cool the magnets to below 2 K has some advantages. Superfluid helium has a very large heat conductivity, which helps to remove heat, and it also has the ability to penetrate the coils through every pore, and therefore to improve the coil cooling. The standard half-cell comprises three dipoles of 14.3 m and one main quadrupole of 3.1 m. Two sets of combined corrector magnets are also installed in each half-cell: a combined octupole/quadrupole corrector and a combined decapole/sextupole/dipole corrector. The R&D program for the LHC has focused on the fabrication of numerous models and prototypes.

The main characteristics of the LHC collider dipoles can be found in Table 3.

**Table 3:** Main parameters of the LHC main dipole and quadrupole magnets

|  | **Dipole** | **Quadrupole** |
|---|---|---|
| Operating temperature (K) | 1.9 | 1.9 |
| Central field (T) | 8.33 |  |
| Gradient (T·m$^{-1}$) |  | 222 |
| Operating current (A) | 11 850 | 11 850 |
| Field length (m) | 14.3 | 3.1 |
| Inner coil diameter (mm) | 56 | 56 |
| Distance between apertures (mm) | 194 | 194 |
| Number of coil layers | 2 | 2 |
| Superconducting material | Nb–Ti | Nb–Ti |
| Collar material | Stainless steel | Stainless steel |
| Yoke | Cold | Cold |
| Total number | 1232 | 386 |

The dipole coils [11] are made with two different cables, one for each layer. The inner layer has 28 strands of 1.065 mm diameter with 7 $\mu$m filaments. The cross-section is 1.736/2.064 × 15.1 mm$^2$. The Cu/Sc ratio is 1.7. The outer layer has 36 strands of 0.825 mm diameter with 6 $\mu$m filaments. The cross-section is 1.362/1.598 × 15.1 mm$^2$. The Cu/Sc ratio is 1.9. The main quadrupole cable is of the same dimensions as that used for the outer layer of the dipole coils. The critical current density at 10 T and 1.8 K in all these cables is above 1530 A·mm$^{-2}$.

The design of the coils is classical with two layers of cables. The Cu/Sc ratio and the cross-section of the cables are adapted on the inner and outer layers to optimize the current density. The insulation scheme is composed of half-overlapped polyimide tape, reinforced with a B-stage polyimide adhesive film.

The force containment consists of coil clamping collars, the iron yoke, and the shrinking cylinder. They contribute to produce the necessary azimuthal pre-compression in the coils and to withstand the electromagnetic forces in one single structure for the two channels. The collars are made of high-strength austenitic steel. They are locked by rods near the mid-plane. The collars are common for the two apertures.

The iron yoke is split vertically into two parts with a gap at room temperature. The gap is needed to compensate for the difference in thermal contraction of the iron and of the collar material during the cooling. The gap is closed at 2 K. The electromagnetic forces at full excitation are shared between the collars and the yoke, and the gap remains closed.

The shrinking cylinder is also the outer shell of the helium tank. This cylinder is welded around the yoke with an interference to generate a tensile stress, which participates in the pre-compression of the coils throughout the current excitation.

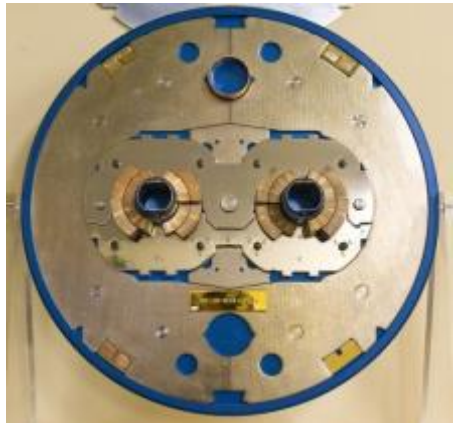

**Fig. 6:** Cross-section of an LHC dipole

Table 3 presents the main characteristics of the lattice quadrupole [12, 13]. Figure 7 shows a cross-section of the cold mass of the quadrupole. The quadrupole design is based on two-in-one geometry. The two quadrupoles of each unit are combined in a focusing/defocusing configuration, to minimize the saturation effects in the yoke. Stainless steel collars alone withstand the electromagnetic forces and the pre-compression. A set of eight tapered keys locks the collars in position. The yoke is made of single-piece laminations, interlocked by pins and keys, with the two quadrupole assemblies fixing their relative positions. A rigid stainless steel tube (inertia tube) is placed around the outside of the yoke and aligns the magnet. Holes, through which dowels can travel, are machined very precisely in the inertia tube. These dowels also pass through the keys, which are placed into slots in the yoke, such that the yoke is aligned by the inertia tube.

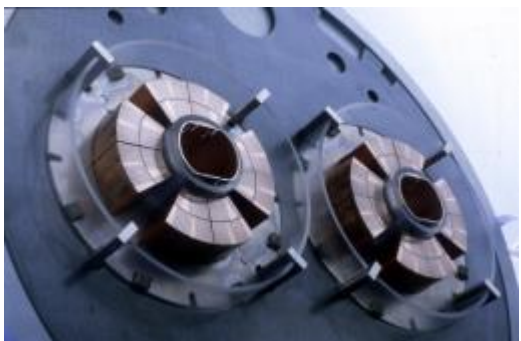

**Fig. 7:** Cross-section of an LHC quadrupole

Most of the magnets reached the nominal current of 11 850 A (about 86% of the short sample limit) with two quenches. Series production magnets have been built by four European companies: Alstom (France), Babcock Noell (Germany), and ASG (Italy) for the dipoles; and ACCEL Instruments (Germany) for the quadrupoles.

## 3 Detector magnets: from the first bubble chambers to CMS and ATLAS

### 3.1 Introduction

Large colliders have emerged in which the experiments can only be carried out at intersection points. The number of detectors installed in the collider is limited by the number of possible intersection points. Therefore, the detector must be designed to carry out the widest range of physics experiments for all the possible directions around the colliding beams. For this purpose, an axially symmetrical configuration is advantageous, and naturally the use of solenoid magnets was selected first. This constraint has led to a new class of magnets for detectors called 'large thin-wall solenoids' [14]. These solenoids provide a huge magnetic field volume (1–4 T) surrounding the colliding beam, filled with a panoply of all-purpose detectors for the analysis of collision events.

A solenoid producing a field in the direction of the circulating beams does not create first-order perturbations on the particles' trajectories and does not require complicated compensating devices. From a technological point of view, a solenoid presents the simplest structure, and, in view of its two-fold symmetry, it is naturally self-supporting and thus raises no difficult mechanical problems. This configuration is particularly favourable in the case of a superconducting solenoid, because it does not require sophisticated cold supports and it allows a simple cryostat structure. The size of the solenoids can be increased to very large dimensions to allow a large volume of free space in which to insert the detectors into the magnetic field. Dipole or toroidal configurations are also used in some detectors. The toroidal field shape gives the best momentum resolution at low angles, and the magnetic field is always transverse to the particle momentum.

Following the above remarks, it is necessary to stress a few requirements specific to the application considered here. The question of transparency is related to the types of particles liable to cross the magnet and reach the detectors situated outside. In practice, two situations can be found according to the relative location of the solenoid with respect to the e–γ calorimeter, which is one of the essential components of the detectors. If the solenoid is inside this calorimeter, it is relatively small in dimension, but it must be made as thin as possible to minimize interactions with particles (typically a fraction of a radiation length, corresponding to a few centimetres of aluminium). If it is outside, it becomes very large in size, but it is not so restricted in material thickness—within the limits imposed by the absorption of hadron particles. Thus, a choice has to be made between the two options, which results from the difficult compromise between considerations of the physics, technical feasibility, and overall cost.

The minimization of the radial thickness of the magnet is also important to save space for the detectors. Other requirements concern the mechanical interfaces in a crowded environment, the weight within the limitation of handling and lifting equipment, and also the cost and reliability.

Generally, an indirect cooling or forced-flow solution has been adopted to keep the temperature of the magnet down. A pool-boiling technique, ensuring cryostatic stability, requires huge and costly helium vessels, incompatible with the requirements of geometry and space. Forced flow in the conductor is sometimes adopted, but this leads to very complex cryogenic systems and serves only to provide an additional reserve of enthalpy for the stability of the conductor. In indirect cooling, loops with helium flowing within are attached to the outer cylinder of the coil. The heat is removed by conduction from the coil to these loops.

The means by which the circulation of the coolant is obtained has also evolved. First, forced-flow modes used the refrigerator Joule–Thompson loop, with the inconvenience of relying completely on the refrigerator. Any shutdown of the refrigerator, however short, leads to a shutdown of the field and often to a quench. Next, forced flow by means of cold pumps with a reserve of helium was introduced. In this case, one depends much less on the refrigerator, but still on the reliability of the pumps. Some more recent projects use natural convection with a 'thermo-siphon' method, working with a reserve of helium placed at the top of the cryostat. This last method has the advantage of being largely independent of the refrigerator and also of not having any moving parts, the reliability of which is always limited.

In the coil design, the first constraint is to ensure safe and reliable operation. The conductor is chosen to have a large safety margin, where the working current is often half the critical current. The conductor is generally a Rutherford cable embedded in a pure aluminium stabilizer. The amount of stabilizer is determined mainly based on criteria for protection, and the overall critical current ranges from 30 to 60 $A \cdot mm^{-2}$, which is about one order of magnitude lower than that currently used for accelerator magnets. Almost all the superconducting solenoids for detectors use the inner winding method, where the first layer of the conductor is layer-wound inside an outer support cylinder. The other layers of the conductors are then wound inside the last layer. The coils are tightly clamped, and their support member is impregnated with epoxy, to form a monolithic structure that ensures good thermal conduction and prevents internal movement of the conductors.

The large dimensions of these solenoids lead to huge hoop stresses that cannot be borne by the conductors alone. For a given field and current density, the hoop stress is proportional to the radius of the solenoid. Therefore, an external support cylinder has to be used at the outer edge of the conductor of the outer layer to limit the tensile strain in the conductor to a value smaller than 0.1%. This cylinder is made of an aluminium alloy with high yield strength. Some conductors are also reinforced with an aluminium alloy outer shell bonded to the pure aluminium stabilizer, to take a part of the large hoop stress developed in some designs.

The stability of a coil is its capacity to absorb or remove possible thermal disturbances, such as conductor motion and deposition of energy by particles. However, such disturbances are difficult to predict as they depend greatly on the quality of the coil's manufacture. We can design a conductor with a very large stability margin, fully cryostable, but which is detrimental to the size of the magnet. Another approach is to limit the large mechanical disturbances by using an epoxy impregnated coil. The design must also avoid local stress concentrations and maintain a stress level far below the initiation of cracks in the epoxy. The conductor itself helps the stabilization with its enthalpy and the high thermal conductivity of the aluminium.

Quench protection is the major requirement in the design of detector magnets, because the stored energy can be as high as 1–3 GJ. During a quench, a large amount of the stored energy is extracted during the dumping of the current in the external resistances with a maximum voltage generally smaller than 1 kV.

At the end of the quench, the maximum temperature reached in the hottest point of the coil must be limited to 100 K, because below this temperature the thermal expansion of the materials is very low and the increase in temperature does not lead to mechanical problems. The Joule heating in the transited length of the conductor is absorbed consistently by the enthalpy of the monolithic conductor itself. This temperature is a function of the dump time constant and the characteristics of the materials. However, despite its higher resistivity, the support cylinder plays a positive role in the protection. This tube is strongly magnetically coupled with the solenoid, such that, when external discharge begins, a part of the energy is transferred to the tube with eddy currents generated by the high d$B$/d$t$. The tube is then heated as a whole, which helps the quench to propagate through the rest of the solenoid ('quench back').

### 3.1.1 Past detector magnets

The first 12 ft superconducting bubble chamber magnet was constructed at Argonne National Laboratory in 1968 [15]. The magnet was a split pair solenoid with a vertical axis and gap to allow the beam entry. Bath-cooled coils with Nb–Ti copper stabilized conductors operated in cryostable mode. A 15 ft version (Fig. 8, left) was built at Fermilab in 1972 [16].

In 1971 at CERN, the Big European Bubble Chamber (BEBC) was built [17], which included a superconducting magnet (Fig. 8, right) made with 20 pancakes of Nb–Ti conductor (3 × 61 mm with 200 untwisted filaments of 200 $\mu$m). The nominal current was 5700 A. The inter-layer insulation included slots to allow boiling He to flow between the conductors. Table 4 gives the main parameters of the first bubble chamber superconducting magnets.

**Table 4:** First bubble chamber superconducting magnets

|  | **Argonne 12 ft** | **Fermilab 15 ft** | **CERN BEBC** |
|---|---|---|---|
| Field (T) | 1.8 | 3 | 3.5 |
| Winding inner diameter (m) | 4.8 | 4.27 | 4.7 |
| Stored energy (MJ) | 80 | 400 | 800 |
| Conductor dimensions (mm) | 2.54 × 50 | 3.8 × 38 | 3 × 61 |
| Current (A) | 2000 | 5000 | 5700 |
| Number of pancakes | 30 | 43 | 20 |

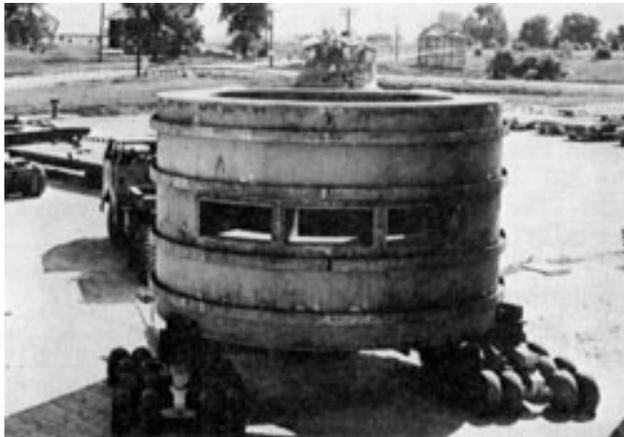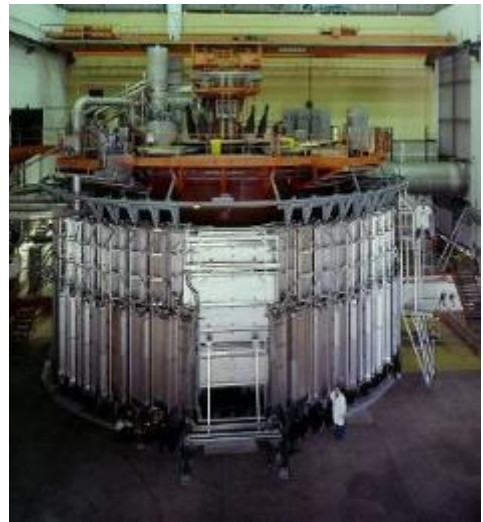

**Fig. 8:** Fermilab 15 ft bubble chamber (left) and BEBC (right)

By the end of the 1970s and during the 1980s, a new type of solenoid for the detectors was developed. To minimize the amount of matter in the coil and its cryostat, low-mass materials were substituted for the usual materials (e.g., an aluminium-stabilized conductor instead of copper and stainless steel). The current density in the conductor was increased (no more adiabatic stability), indirect cooling by external pipes was implemented, and an intrinsic protection with aluminium shunts and quench back tube was designed to allow a quasi-uniform distribution of the stored energy in case of quench.

A list of such solenoids and their main characteristics is given in Table 5.

DELPHI [18] and ALEPH [19] were the two detectors for the LEP using superconducting coils. They both had to produce a very uniform field in the large volume of the central detector, and for that purpose they included additional compensating windings at the two ends of the main solenoid. Both use Al-stabilized conductors. A new feature for this type of conductor compared with the older ones is the use of flat Rutherford cable as an insert instead of a monolithic superconductor.

**Table 5:** Characteristics of the last generation of superconducting solenoids in operation today

| Name | ALEPH | DELPHI | CLEO2 [20] | ZEUS [21] | H1 [22] |
|---|---|---|---|---|---|
| Accelerator | LEP | LEP | CESR | HERA | HERA |
| Laboratory | CERN | CERN | Cornell | DESY | DESY |
| Designed by | Saclay | RAL | Cornell Oxford. | Milan University. | RAL |
| Manufactured by | Saclay | RAL | Oxford Inst. | Ansaldo | RAL |
| Inner bore (m) | 4.96 | 5.2 | 2.88 | 1.72 | 5.2 |
| Outer bore (m) | 5.98 | 6.2 | 3.48 | 2.22 | 6.08 |
| Winding length (m) | 6.35 | 6.8 | 3.48 | 2.5 | 5.16 |
| Overall length (m) | 7 | 7.4 | 3.78 | 2.8 | 5.75 |
| Conductor (mm$^2$) | 35 × 3.6 | 24 × 4.5 | 16 × 5 | 15 × 4.3 | 26 × 4.5 |
| Stabilizer | Al | Al | Al | Al | Al |
| Cold mass (t) | 25 | 25 | 7 | 3.4 | |
| Conductor mass (t | 8 | 7 | | 2.4 | 7 |
| Current (A) | 5000 | 5000 | 3300 | 5000 | 5500 |
| Design field (T) | 1.5 | 1.2 | 1.5 | 1.8 | 1.2 |
| Stored energy (MJ) | 137 | 108 | 25 | 16 | 120 |
| Cooling method | Thermo-siphon | Forced-flow pumps | Thermo-siphon | Forced-flow pumps | Forced-flow pumps |
| Radiation length ($X_0$) | 2 | 7.4 | | 0.9 | 4 |
| Year of completion | 1987 | 1988 | 1988 | 1989 | 1989 |

The ALEPH winding is made in two halves, each of 3.2 m, whereas the DELPHI coil is split into shorter modules of 1.5 m length. The impregnation technique is also different: DELPHI uses a 'prepreg' insulated conductor and ALEPH is vacuum impregnated. The two solenoids use the 'inner winding' method, in which the conductor is layered from an external spool on the inner radius of the external support cylinder through a number of fixtures. The cooling circuit is laid in a similar way for both magnets with loops on the support cylinder. However, the refrigeration systems are different. For DELPHI, liquid helium is circulated in forced flow with cold pumps, whereas ALEPH uses a thermo-siphon.

## 3.2 Detector for LHC projects

The energy stored in the detector magnets for the LHC is one order of magnitude larger than in previous generations of magnets because of their size and magnetic field. They all use a cable embedded in aluminium with indirect cooling as the conductor. All solenoids use an iron yoke. The final problem is that the magnets must be assembled on site because of their size.

### 3.2.1 The CMS detector for LHC

The Compact Muon Solenoid (CMS) [22] (Fig. 9, left) is a high-field superconducting magnet (4 T) with a massive return yoke (12 000 ton) (Fig. 9, right), which forms part of the external muon spectrometer. The level of field is required to achieve a high resolution compatible with the high luminosity of the LHC.

The main characteristics are a nominal magnetic field of 4 T at the operating current of 19.5 kA, in a 5.9 m diameter and 12.5 m long warm bore, leading to a stored energy of 2.7 GJ. These characteristics make this superconducting solenoid the largest and most powerful ever designed. The CMS solenoid has a larger stored energy per unit mass of coil (12 kJ·kg$^{-1}$) than any large superconducting magnet ever built.

The coil comprises a four-layer winding to increase the number of ampere-turns, divided into five identical sections of 2.5 m length, each internally wound and vacuum impregnated before final assembly. The outer part of these modules (a 50 mm thick aluminium alloy shell) is used as an outer winding mandrel and as a mechanical reinforcement structure, but also as a cooling wall and quench back tube during cool-down, energization, and fast discharge of the coil. The cold mass total weight is about 225 t. It is supported inside the vacuum tank through a system of radial and longitudinal tie-bar supports.

The conductor is made of three parts, which are assembled during the fabrication process: a Rutherford cable made of 32 strands of 1.288 mm diameter is co-extruded inside a high-purity aluminium matrix to form what is called the insert; two aluminium alloy sections are then welded on each side of the insert using the continuous electron beam technique to give the mechanical reinforcement to the conductor structure. The critical current of the conductor at 5 T and 4.2 K is 55 600 A. Cable dimensions are 20.63 × 2.34 mm$^2$, and the conductor dimensions are 64 × 22 mm$^2$.

Indirect cooling of the cold mass is performed using a two-phase helium thermo-siphon.

The five modules were delivered over a period of one year up to January 2005 and superposed vertically around the internal vacuum chamber and also supporting the internal thermal shields. Then, after the external thermal shields were put in place, the assembly was tilted into the horizontal position and inserted into the external vacuum chamber, and positioned in the magnetic yoke in August 2005. Following the successful testing of the detector magnet at its nominal field rating of 4 T in the CERN surface facility between August and November 2006, the detector was disassembled to be lowered into the underground cavern. The magnet, together with its central soft iron shielding ring, was lowered into the tunnel in February 2007. The magnet was ramped up to its nominal field level in November 2008, with no serious problems occurring.

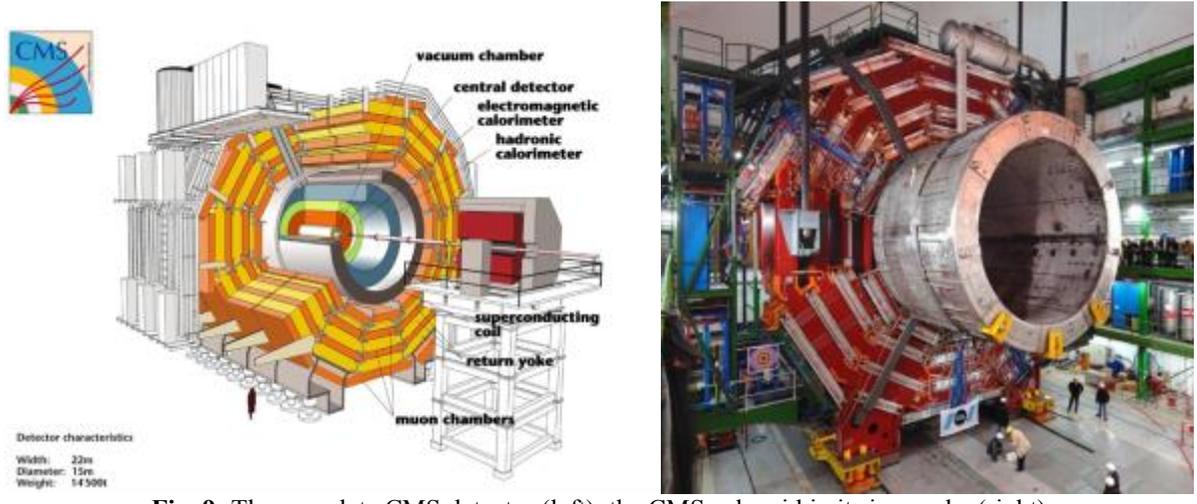

**Fig. 9:** The complete CMS detector (left); the CMS solenoid in its iron yoke (right)

### 3.2.2 The ATLAS detector for LHC

ATLAS is one of the two experiments dedicated to the search for the Higgs boson installed on the LHC ring at CERN.

The ATLAS detector contains two separate magnetic systems [4]: a thin-wall superconducting solenoid close to the central detector and a set of large superconducting toroidal magnets for the outer muon spectrometer (Table 6).

The complete toroidal system comprises three separate toroids: a large central 'barrel' toroid (Fig. 10) and two 'end cap' smaller toroids inserted at each end of the barrel to achieve complete angular coverage. Each toroid is made with eight 'racetrack' coils laid radially around the beam axis. Each coil is built independently and consists of two double pancakes, epoxy impregnated, together with pure aluminium cooling plates, which conduct the heat to the liquid helium loops (classical indirect cooling scheme). A central web structure is clamped rigidly to the two pancakes and contains, via pure compression, the internal magnetic forces applied to the conductors. The conductor is a Rutherford cable co-extruded in a pure aluminium matrix.

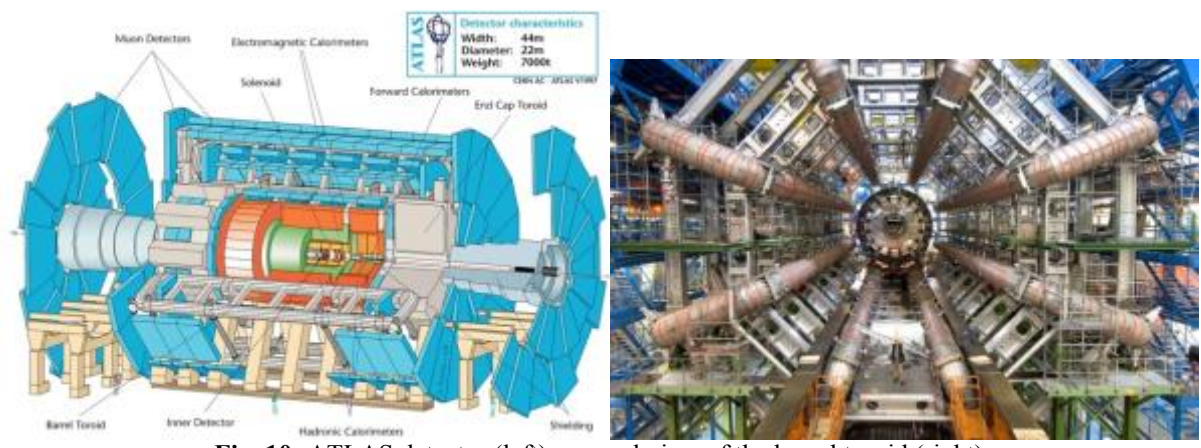

**Fig. 10:** ATLAS detector (left); general view of the barrel toroid (right)

Table 6: Main parameters of the superconducting toroids and solenoid for ATLAS

|  | Barrel | End cap | Solenoid |
|---|---|---|---|
| Inner bore (m) | 9.4 | 1.65 | 2.46 |
| Outer diameter (m) | 20.1 | 10.7 | 2.63 |
| Overall length (m) | 25.3 | 5 | 12.5 |
| Number of coils | 8 | 2 × 8 | 1 |
| Stored energy (MJ) | 1080 | 2 × 250 | 40 |
| Operating current (kA) | 20.5 | 20.5 | 7.7 |
| Peak field (T) | 3.9 | 4.1 | 2.6 |
| Total weight (t) | 830 | 2 × 239 | 5.7 |
| **Conductor** | | | |
| Overall size (mm$^2$) | 57 × 12 | 41 × 12 | 30 × 4.25 |
| Type | Rutherford cable + pure Al co-extruded | | Rutherford cable + high-strength pure Al co-extruded |

The main challenge for these magnets and especially the 'barrel toroid', is their integration in the complex detector set-up. The magnet must leave the maximum open space for the detectors, and therefore the support structure must be as invisible as possible to the particles.

The eight coils for the central 'barrel toroid', developed by CEA Saclay in collaboration with LASA (INFN) and CERN, were tested one at a time, above ground, by applying a 22 000 A current. In 2005, they were assembled in the ATLAS cavern, 100 m below ground level, using an aluminium structure. Arranged in a star configuration and positioned with precision of a few millimetres, the coils occupy a volume equal to that of a six-storey building. The structure supporting the 1400 ton muon detector can withstand considerable magnetic forces.

The 2 T central toroid, which is a magnet developed by the High Energy Accelerator Research Organization (KEK) in Japan, was cooled to 4 K for the first time in June and July 2006. A 21 000 A test current was then applied during the night of 9 November 2006. In view of the success of the first performance test, the central toroid was connected to the end toroids, which were developed by the Rutherford Appleton Laboratory in collaboration with NIKHEF and CERN. The two end toroids are drawn to the central toroid by a force of 240 t.

The three magnets were brought up to their rated current on 4 August 2008. The toroids were then tested at the same time as the central solenoid.

All these performance tests were successful, and ATLAS is now the largest superconducting magnet system in the world.

# 4 Fusion magnets: review of machines built and under construction

## 4.1 Introduction

The thermonuclear plasmas produced in fusion machines need to be confined by strong magnetic fields. In the early days, resistive magnets were used in small-sized machines and in pulse-mode operations. However, the total electrical power needed to energize the magnets was sometimes greater than 1 GW, which drove the need for machines using superconducting magnets.

The development of superconducting magnets for fusion started in the 1970s, and today all large fusion projects present a superconducting magnet system and no longer use resistive magnets. A detailed description of the history of fusion machines can be found in Ref. [23].

## 4.2 Baseball magnets

The Baseball I and II (Fig. 11) superconducting magnets were developed in the 1960s for magnetic fusion experiments at Lawrence Livermore National Laboratory [24]. The Baseball II superconducting magnet was designed to generate a maximum magnetic field of 7.5 T (for a central field strength of 2 T). The designed nominal current was 2400 A. The conductor was a monolith of 6.35 mm$^2$, made with a composite of 24 Nb–Ti 0.6 mm filaments embedded in a copper matrix. The magnet weighed 13 t and has been operated at central fields of up to 1.5 T.

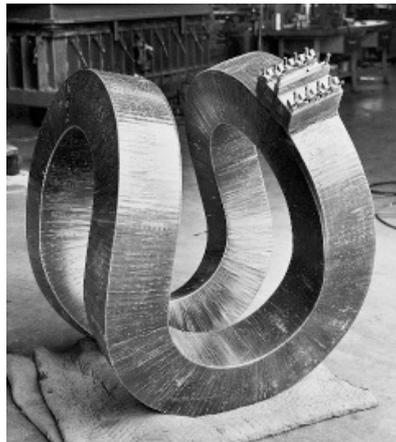

**Fig. 11:** Baseball II superconducting coil

## 4.3 The Mirror Fusion Test Facility

The Mirror Fusion Test Facility (MFTF), built at Lawrence Livermore National Laboratory, is an advanced experimental fusion device designed as an intermediate step between the existing mirror machines and an experimental fusion reactor. MFTF was completed in 1986, but the overall project was cancelled later that year. The MFTF magnet system [25] contained a total of 26 large superconducting coils made with yin-yang pairs and solenoids.

The yin-yang magnet pair (Fig. 12, left) had an average major radius of 2.5 m and an average minor radius of 0.75 m. A peak magnetic field of 7.68 T occurred on the windings at the minor radius. The magnetic field dropped rapidly to 4.2 T at the mirrors and to 2.0 T in the centre.

The first yin-yang magnet was tested successfully in February 1982 to its full design field (7.68 T) and current (5775 A). The coil weighed 341 ton, and the stored energy was 410 MJ. The coil current density was 25 A·mm$^{-2}$. The 480 Nb–Ti 0.2 mm diameter filaments were embedded in a copper matrix (Cu/Sc ratio of 1.7). The conductor (Fig. 12, right) was wrapped and soldered in an embossed and perforated sheath of high-purity copper to improve the stability and the cooling. The overall dimensions were 12.4 × 12.4 mm. The coil was wound with 58 layers of 24-turn conductors.

The magnet was cooled in a 4.3 K, 1.3 bar helium bath. The magnet weighted 375 ton for a stored energy of 410 MJ.

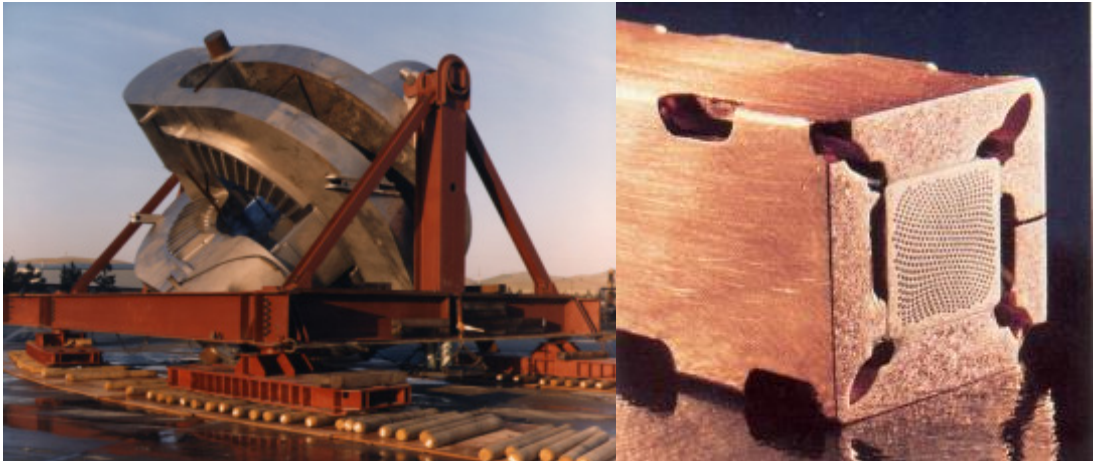

**Fig. 12:** MFTF yin-yang coils (left) and conductor (right)

**4.4   T-15 Tokamak**

The Tokamak T-15 [26, 27] was built in the Kurchatov Institute, Moscow, Russia, at about the same time (1983–1988) as the Tore Supra in Cadarache (see Section 4.5), using 24 TF coils (Fig. 13, left) with over 11 ton of $Nb_3Sn$ material. It was the largest $Nb_3Sn$-based device in the world prior to the start of construction on the ITER.

The coils were wound with a react and wind conductor made with a flat cable of 10 $Nb_3Sn$ 1.5 mm strands, first heat treated on a drum and then temporarily straightened to be bonded to two copper pipes of 0.4 m diameter by an electroplated Cu layer (Fig. 13, right). The critical current of the 100 km conductor was 3.9 kA at 6.5 T. The overall dimensions were 17.4 × 6.5 mm. T-15 produced its first plasma in 1988, and it was shut down in 1995 due to the poor economic environment in Russia at that time. The TF coil system was made of 24 circular coils with an average diameter of 2.4 m and a stored energy of 384 MJ.

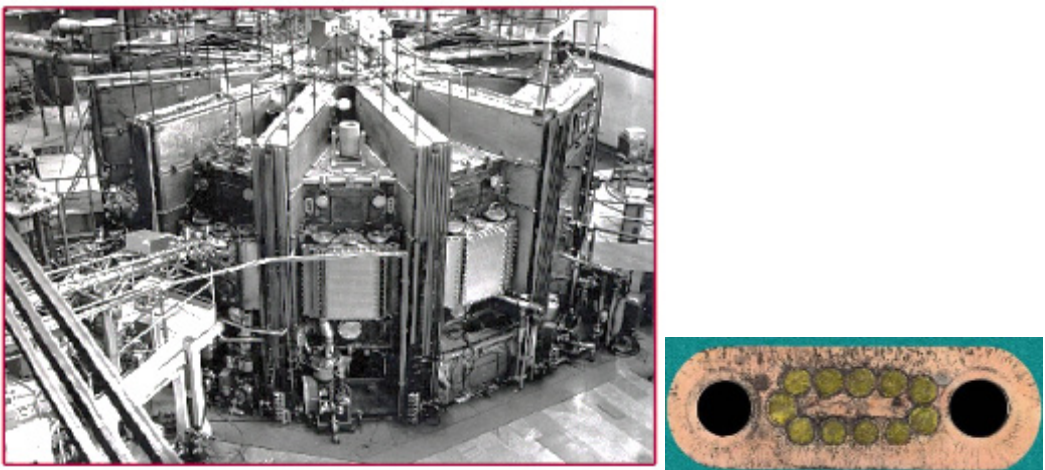

**Fig. 13:** T-15 TF coils (left) and conductor (right)

**4.5   Tore Supra**

Tore Supra is a superconducting Tokamak installed at the CEA Cadarache by the Euratom-CEA Association. It delivered its first plasma in April 1988. The major plasma radius is 2.25 m and the minor radius is 0.7 m. The overall diameter is 11.5 m and the height is 7.5 m.

Each Toroidal Field (TF) coil [28] (Fig. 14, left) is made of 26 double pancakes of conductor separated by insulating spacers, which provide both the mechanical cohesion of the winding and secure a very large volume of superfluid helium at 1.8 K and 1 bar in direct contact with the conductor. Each coil winding is encased in a strong stainless steel casing. The 18 cases form an inside and an outside vault, which resists the centripetal forces produced by the toroidal field.

The Nb–Ti conductor (Fig. 14, right) is a monolithic bare conductor with 23 $\mu$m filaments embedded in a copper matrix. The dimensions are 2.8 × 5.6 mm. The weight of the superconductor is about 45 t for a total magnet weight of 160 t. The stored energy is 610 MJ.

Full performances of the magnet were reached on 8 November 1989. The current in the TF coils was increased to 1455 A corresponding to 9.3 T on the conductor (design values: 1400 A and 9 T, respectively) and 4.5 T at the plasma centre. Tore Supra coils are now operated at 90% of the design values (4 T on the plasma). Tore Supra demonstrates that a TF superconducting system can be operated routinely in the severe conditions of a Tokamak.

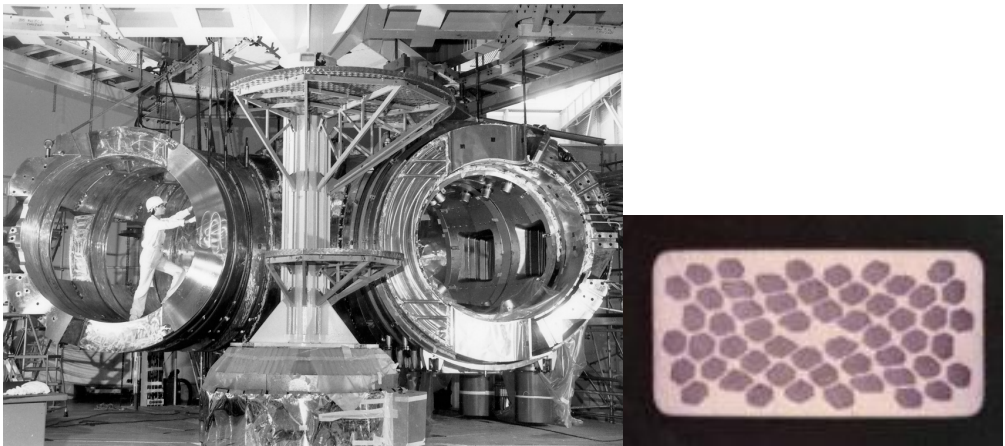

**Fig. 14:** Tore Supra TF coils (left) and conductor (right)

### 4.6 Experimental Advanced Superconducting Tokamak (EAST)

EAST is the Experimental Advanced Superconducting Tokamak built at the Institute of Plasma Physics of the Chinese Academy of Sciences (CASIPP) in Hefei [29]. EAST's major radius is 1.7 m and its minor radius is 0.4 m. The machine's overall dimensions are 10 m height and 7.6 m diameter. The total weight is 414 ton. The toroidal field at the plasma centre is 3.5 T, the operating current is 14.3 kA, and the total stored energy is 300 MJ.

The magnetic system comprises 16 D-shaped TF coils (Fig. 15), one central solenoid split into 6 individual coils, and 4 large Poloidal Field (PF) coils. The peak field on the TF coil is 5.8 T. The largest PF coil has an outer diameter of 7.6 m. Coils are cooled with supercritical helium at 4.5 K, and the total cold mass of the superconducting magnet system is 194 ton.

The TF conductor is an Nb–Ti/Cu Cable in Conduit Conductor CICC. The diameter of the superconducting strands is 0.85 mm. The temperature margin is 1.88 K. The jacket material of the cable is 316LN stainless steel with 1.5 mm wall thickness. The dimensions of the CICC are 20.4 × 20.4 mm. The turn electrical insulation consists of multiple layers of polyimide and fibreglass tapes impregnated with epoxy resin.

The assembly of the Tokamak was completed at the end of 2005, and the first engineering commissioning was carried out at the beginning of 2006. The EAST machine is now in operation.

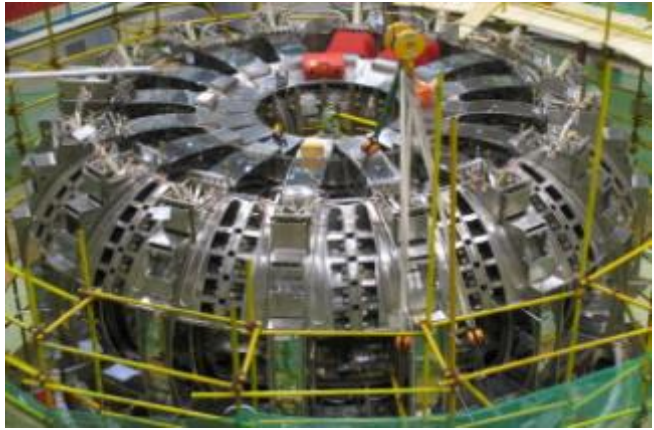

**Fig. 15:** EAST TF coil assembly

### 4.7 SST-1

SST-1 (Steady State Superconducting Tokamak) is a plasma confinement experimental device built at the Institute for Plasma Research in Gandhinagar, India. The project started in 1994 and the integration of the system was completed in 2012. The first plasma campaigns have already started, following engineering validation of the Tokamak.

The magnet System of SST-1 [30] is an assembly of 16 superconducting D-shaped TF coils (Fig. 16) with an average diameter of 1.8 m, 9 superconducting PF coils, and a pair of resistive PF coils. TF magnets generate a 3.0 T field at the major radius of 1.1 m. The minor plasma radius is 0.2 m. The diameter of the machine is 4.4 m and its height is 2.6 m. The total weight is 160 ton. Magnets are cooled with supercritical helium at 4 bar and 4.5 K. The operating current is 10 kA and the peak field on the coils is 5.1 T. Each of the TF coils consists of six double pancakes, each pancake having nine turns. The TF system stores 56 MJ.

The conductor is an Nb–Ti CICC. The PF superconducting magnets are wound from the same CICC as that used for the TF magnets. The conductor consists of 1350.86 mm diameter NbTi/Cu strands with a high copper to superconductor ratio of 4.9:1. These strands are twisted in four stages before being jacketed inside a conduit made of stainless steel, which has a cross-section of 14.8 × 14.8 mm and a void fraction of 40% inside the cable space for liquid helium to flow.

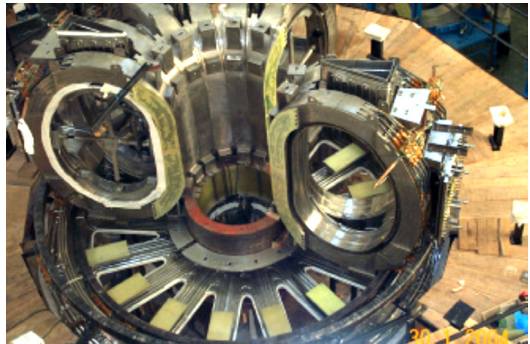

**Fig. 16:** SST-1 TF coils

### 4.8 KSTAR

The Korean Superconducting Tokamak Advanced Reactor (KSTAR) (Fig. 17, left) is an advanced plasma, steady-state Tokamak experiment built at the Korean Basic Science Institute in Daejon, South Korea [31]. The system has been in operation since 2008.

The 16 TF coils (Fig. 17, right) provide a 3.5 T magnetic field at the centre of the plasma. The major radius is 1.8 m, for a minor radius of 0.5 m. The TF coil dimensions are 4.2 m height and 3 m width. The overall dimensions of the machine are 8.8 m diameter and 8.6 m height. The peak field of the TF coils is 7.2 T. The PF system has 13 coils, 7 in a Central Solenoid (CS) stack and 6 outer PF coils.

All conductors use CIC superconductors with cooling by forced-flow supercritical helium with an inlet temperature of 4.5 K and an inlet pressure of 5 bar. The TF coils use 0.78 mm diameter $Nb_3Sn$ strands in a 2.8 mm thick Incoloy 908 conduit. Conductor dimensions are 25.65 × 25.65 mm. The cable pattern is 34 × 6 of 486 strands. The conductor current in the TF coils is 35.2 kA and the stored energy is 470 MJ. The critical current density of the $Nb_3Sn$ strands at 12 T and 4.2 K is higher than 750 A·$mm^{-2}$. CS and PF coils use $Nb_3Sn$ in an Incoloy 908 conduit, except for the outer PF coils that use NbTi strands in a 316LN conduit.

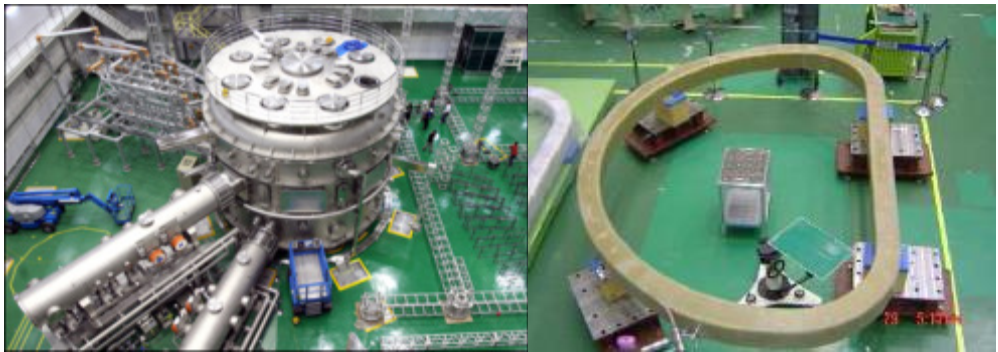

**Fig. 17:** KSTAR machine (left); TF coil (right)

**4.9  JT-60SA**

The magnet system for the JT-60SA [32] is developed within the framework of the ITER 'Broader Approach'. An agreement has been signed between Japan and Europe to upgrade the Japanese JT-60 Tokamak to a superconducting Tokamak, JT-60SA (Fig. 18), as a 'satellite' facility to ITER, to develop operating scenarios and address key physics issues for an efficient start of the ITER experiment and for research towards the future DEMOnstration Power Plant for Fusion (DEMO). Europe is in charge of supplying TF coils for plasma confinement in the machine, as well as other components (power supplies, current leads, cryogenic plant, ECRH, and cryostat), as an in-kind contribution to the project.

The major plasma radius is 3.1 m and the minor plasma radius is 1.15 m. The toroidal magnetic field at the plasma centre is 2.68 T for a peak field on the conductor of 6.5 T. The operating current is 25.7 kA and the temperature margin is 4.6 K. The total mass of the magnet is about 1300 t and the total magnetic energy is 1060 MJ.

Eighteen TF coils, a CS, and seven Equilibrium Field (EF) coils are at the heart of the Tokamak system. The four CS modules use $Nb_3Sn$-type superconductors, whereas the TF and EF coils are made of Nb−Ti. All conductors for the TF, CS, and EF coils are cooled with supercritical helium with a coil inlet temperature of 4.5 K. The TF coils use NbTi superconductor at 5.65 T. The TF conductor is a CICC with a circular multistage cable comprising 486 strands cabled without a central spiral. The minimum temperature margin is 1.2 K under normal operating conditions and 1.0 K after a plasma disruption. The operating current is 25.7 kA for the TF coils. The TF conductor has an outer dimension of 22 × 26 mm, which is deliberately not square in order to optimize the winding pack (six double pancakes with six turns) shape in the TF coil case. The CS operates at high field and uses $Nb_3Sn$ superconductor.

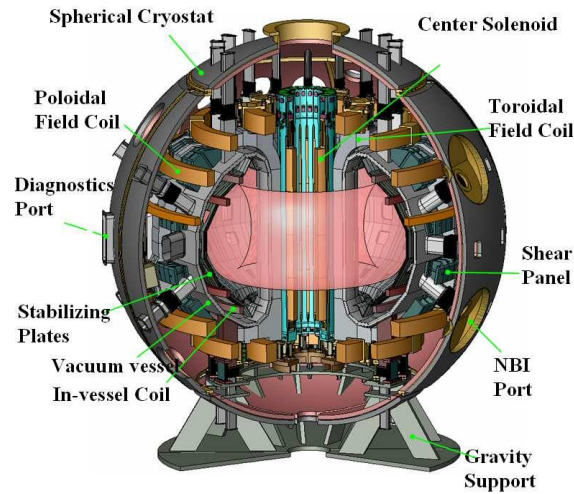

**Fig. 18:** JT-60SA Tokamak

### 4.10 Wendelstein 7-X

Stellarators use a single-coil system with no longitudinal net current in the plasma, and hence operate without a transformer (continuous operation and inherent stability). The Wendelstein 7-X (W7-X) machine is a low-shear stellarator under construction at IPP Greifswald, Germany [33, 34]. The major radius is 5.5 m and the minor radius is 0.53 m. The magnetic field on the axis is 3 T and the stored energy is 600 MJ. The machine has a mass of 725 t. The assembly of the W7-X magnets (Fig. 19, left) are at an advanced stage, and the first plasma is scheduled for 2015.

The magnet system is formed by 70 superconducting coils: 50 Non-Planar Coils (NPC) (Fig. 19, right) with 5 different geometries are wound from 960 m of Nb−Ti conductor each, and 20 Planar Coils (PC) with 2 different geometries are wound from 390 m of conductor each. The coil windings are manufactured with a superconducting Nb−Ti cable enclosed in an aluminium jacket. After vacuum pressure impregnation, the windings are embedded in cast stainless steel casings, which are then equipped with cooling systems consisting of copper strips or plates soldered to stainless steel tubes. Following factory tests, all coils are tested at cryogenic temperature in a facility at CEA Saclay [35] as part of the final and formal acceptance procedure.

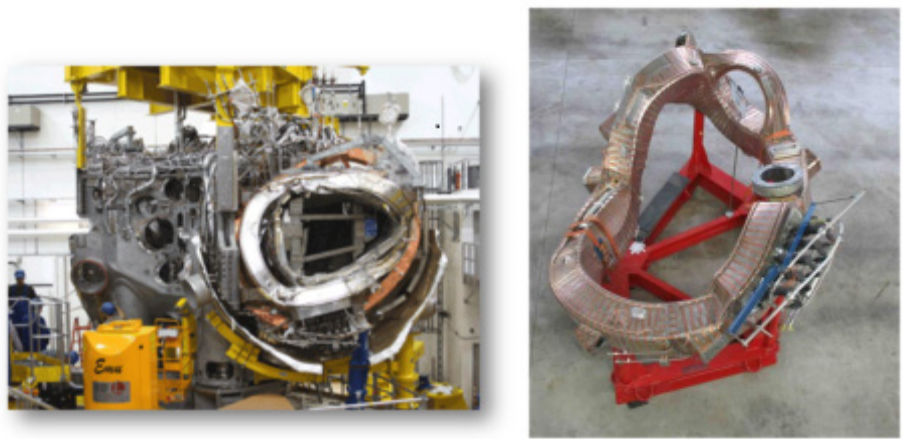

**Fig. 19**: W7X machine assembly (left); NPC (right)

## 4.11 ITER

ITER (Fig. 20, left), the International Thermonuclear Experimental Reactor Programme, should demonstrate the scientific and technological feasibility of fusion power by achieving extended burn of D–T plasmas with steady state as the ultimate goal. The reactor is being built at Cadarache near Aix-en-Provence in France, and the first plasma is planned for 2020. The other programme goals are to integrate and test all essential fusion power reactor technologies and components, and to demonstrate safety and environmental acceptability of fusion for future machines. The major radius of the machine is 6.2 m and the minor radius is 2 m. The magnetic field at the centre of the plasma is 5.3 T.

The ITER magnetic field is composed of four systems [5]: the toroidal magnetic field system, the CS, the PF system, and the Correction Coils (CC). They all use Nb–Ti- and $Nb_3Sn$-based conductors. The total mass of the system is about 10 000 t. Table 7 lists the main parameters of the coils.

The conductors are CICC made up of superconducting and copper strands assembled into a multistage, rope-type cable inserted into a conduit of butt-welded austenitic steel tubes (Fig. 20, right). Table 8 gives the main parameters of the ITER conductors.

The coils are cooled with supercritical helium at an inlet temperature of 4.5 K.

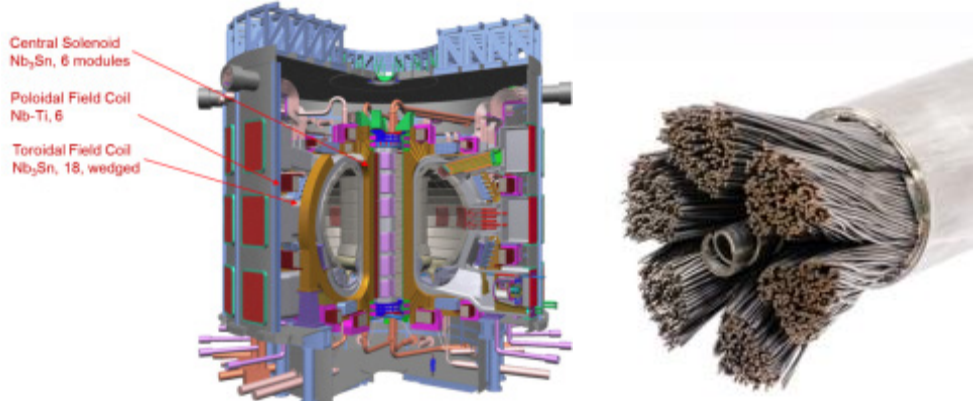

**Fig. 20**: ITER Reactor (left); TF conductor (right)

**Table 7:** ITER coil characteristics

|  | TF | CS | PF |
|---|---|---|---|
| Number of coils | 18 | 1 (6 modules) | 6 |
| Dimensions (m) | 14 × 9 | 12 × 4 | 8–24 |
| Conductor type | $Nb_3Sn$ CICC | $Nb_3Sn$ CIC | $Nb_3Sn$ CIC |
| Quantity (km) | 88 | 42 | 65 |
| Conductor total weight (t) | 826 | 728 | 1224 |
| SC strand weight (t) | 384 | 122 | 224 |
| Operating current (kA) | 68 | 45 | 45 |
| Operating temperature (K) | 5 | 4.5 | 4.5 |
| Peak field (T) | 11.8 | 12.8 | 6 (on PF6) |
| Stored energy (GJ) | 41 | 7 | 4 |
| Coil total weight (t) | 6540 | 974 | 2163 |

Table 8: ITER conductor parameters

|  | TF | CS | PF |
|---|---|---|---|
| Conductor | $Nb_3Sn$ | $Nb_3Sn$ | Nb–Ti |
| CIC configuration | 226 × 6 | 226 × 6 | 240 × 6 |
| Conductor dimension (mm) | 43.7 | 49 | |
| Number of superconducting strands | 900 | 576 | 1440 |
| Number of copper strands | 522 | 288 | |
| Diameter of superconducting strands (mm) | 0.82 | 0.83 | 0.73 |
| Jacket material | 316LN | JK2LB | 316LN |
| Critical current/strand (A) @12 T, 4.2 K | 190 | 220 | |
| Critical current/strand (A) @5.6 T, 4.2 K | | | 303 |
| Total mass (t) | 826 | 745 | 1224 |
| Operating temperature margin (K) | 0.8 | 0.8 | 1.6 |
| Helium fraction (%) | 29.7 | 33.5 | 34.2 |

## 5   Conclusion

Important developments have been made in the technology over the last 40 years for the large-scale applications of superconductivity in terms of field strength, scale, field volume, and stored energy.

The development of new conductors has increased their capabilities to withstand high current densities and large mechanical forces and stresses. New coil winding configurations and new coil assembly methods enable the engineering of large magnets with sizes and stored energies never achieved before.

The next step will be to use $Nb_3Sn$ and HTS materials to increase the magnetic field level and magnetized volumes. Robust R&D is needed to enhance the conductor mechanical strength and to protect the coils against quenches, which will provide an opportunity for the next generation of large-scale superconducting magnets to emerge.

**Acknowledgements**

I wish to thank all those people who have contributed to this paper, and particularly: Jean-Luc Duchateau (CEA), Arnaud Devred (ITER), François Kircher (CEA), Elwyn Baynham (RAL), Akira Yamamoto (KEK), Lucio Rossi (CERN), Luca Bottura (CERN), and Paolo Ferracin (CERN) for all their contributions of material.